\documentclass[11pt]{article}

\usepackage[margin=1in]{geometry}
\usepackage{amsmath,amssymb}
\usepackage{booktabs,tabularx,array}
\usepackage{graphicx}
\usepackage{xcolor}
\usepackage{microtype}
\usepackage[round,authoryear]{natbib}

\graphicspath{{../figures/}}
\newcolumntype{Y}{>{\raggedright\arraybackslash}X}

\title{\textbf{First-passage tests of anomalous transport closures in heterogeneous aquifers}}
\author{Ke-Ming Shen$^{a,b,*}$ and Yi-Xuan Liu$^{c}$\\[0.4em]
\small $^{a}$School of Science, East China University of Technology, Nanchang 330013, China\\
\small $^{b}$Key Laboratory of Quark and Lepton Physics (MOE), Central China Normal University,\\
\small Wuhan 430079, China\\
\small $^{c}$School of Water Resources and Environmental Engineering, East China University of Technology,\\
\small Nanchang 330013, China\\[0.4em]
\small $^{*}$Corresponding author: \texttt{shen\_keming@ecut.edu.cn}}
\date{}

\begin{document}
\maketitle

\begin{abstract}
Breakthrough curves (BTCs) record arrival times, but anomalous transport models are often fitted with time-normalized spatial concentration profiles. Here, the effect of this mismatch on groundwater-transport upscaling is tested. Analytical limits and high-$q$ scaling were verified first. Six first-passage models were then fitted to particle arrivals from 36 two-dimensional digital aquifer runs. The set included exact ADE, $q$ and time-memory extensions, a joint memory-$q$ model, Gamma-CTRW, and a two-path inverse-Gaussian mixture. After the operator was corrected, memory-qFPT was not selected in any run. AICc selected the two-path model in 20 runs, Gamma-CTRW in 15, and q-FPT in one. Median downstream holdout log-RMSE was 0.904, 0.914, and 0.959 for the two-path, Gamma-CTRW, and memory-qFPT models, respectively. The memory-$q$ parameters also failed an independent plume-scaling test ($R^2=0.189$; slope $=-0.332$). Meanwhile, multimodal downstream BTCs increased from 0\% in homogeneous media to 100\% in the strongest heterogeneity class. Gamma-CTRW was also favored in an independent synthetic field. Observation consistency is therefore necessary but not sufficient for physical validation. In these simulations, pathway and waiting-time structure was supported more strongly than a unique memory-$q$ law.
\end{abstract}


\section{Introduction}

Preferential paths, broad velocity distributions, and local retention are created by hydraulic-conductivity heterogeneity. Early arrival and long BTC tails can then persist over large distances. In such cases, a constant-dispersion model may fail even when ADE fits part of the record \citep{berkowitz2000,berkowitz2006,comolli2019}. Several alternatives have been proposed. They include CTRW, multirate mass transfer, fractional transport, and nonlinear diffusion \citep{haggerty1995,metzler2000,benson2001,hansen2020}. Similar tails can be produced by these different mechanisms. Calibration alone is therefore not enough to identify the governing process.

Non-Gaussian spreading can be represented compactly by a nonlinear Fokker-Planck equation. In one solution family, the q-Gaussian shape and its width scaling are controlled by $q$ \citep{tsallis1996}. Memory can be added through a fractional time derivative of order $\beta$. A four-parameter model is obtained when both effects are combined. However, a physical meaning cannot be assigned from fit quality alone. The fitted $q$ and $\beta$ must predict unused data. An independently measured spreading exponent must also be recovered. When data are generated by the same equation, only internal identifiability is shown.

A more basic condition is set by the observation itself. Particle crossing times at a control plane are first-passage data. By contrast, a concentration propagator $p(x,t)$ gives occupancy in space. A first-passage density is not generally obtained by normalizing $p(x,t)$ over time \citep{naff1992,redner2001,artime2018}. Even for ADE, an $x/t$ flux factor is required. If the wrong operator is used, the error may be absorbed by anomalous parameters. A complex model may then appear to be supported for the wrong reason.

Three questions are addressed. First, how much support for memory q-FPE remains after a first-passage operator is used? Second, can the corrected model pass distance, flow, realization, and scaling tests? Third, are the arrivals explained better by explicit waiting times or multiple pathways? Memory q-FPE is treated as a hypothesis throughout. All laboratory-like and field-like records are simulated. Thus, a mechanism can be tested under controlled conditions, but empirical field validation cannot be claimed.

\section{Theory and hypotheses}

\subsection{Memory nonlinear Fokker-Planck hypothesis}

The candidate spatial closure is
\begin{equation}
{}^{C}D_t^{\beta}p(x,t)
=-v_{\beta}\frac{\partial p}{\partial x}
+D_{q,\beta}\frac{\partial^2 p^{2-q}}{\partial x^2},
\label{eq:mqfpe}
\end{equation}
Here, $p$ is a normalized spatial density. Nonlinear spreading is controlled by $q$, and $\beta\in(0,1]$ is the Caputo order. Generalized drift and dispersion are denoted by $v_{\beta}$ and $D_{q,\beta}$. Four nested cases are obtained: ADE ($q=\beta=1$), q-FPE ($q\ne1,\beta=1$), time-fractional ADE ($q=1,\beta\ne1$), and memory q-FPE ($q\ne1,\beta\ne1$).

For zero drift, a self-similar balance gives the finite-variance spreading hypothesis
\begin{equation}
\sigma_x^2(t)\propto t^{\alpha},\qquad
\alpha_{q,\beta}=\frac{2\beta}{3-q}.
\label{eq:scaling}
\end{equation}
Near $q=5/3$, sample second moments are strongly affected by finite domains and remote tails. Equation~\eqref{eq:scaling} is therefore checked with both mean-square displacement and squared interquartile width (IQR$^2$).

\subsection{First-passage observation models}

For one-dimensional ADE, a solute pulse is released at the origin and an absorbing plane is placed at $x>0$. With drift $v>0$ and dispersion $D$, the exact first-passage density is inverse Gaussian:
\begin{equation}
f_{\mathrm{ADE}}(t\mid x)=
\frac{x}{\sqrt{4\pi D t^3}}
\exp\!\left[-\frac{(x-vt)^2}{4Dt}\right].
\label{eq:ig}
\end{equation}
To match the processed particle records, the sampled density is normalized over the available time window.

No exact absorbing-boundary solution is assumed for Eq.~\eqref{eq:mqfpe}. Instead, the following reduced first-passage model is used to test operator sensitivity:
\begin{equation}
f_{q,\beta}(t\mid x)=\mathcal{N}_{t}^{-1}
\frac{x}{t}\,G_q\!\left[x;
\frac{v t^{\beta}}{\Gamma(1+\beta)},
2D t^{2\beta/(3-q)}\right],
\label{eq:qfpt}
\end{equation}
Here, $G_q$ is a finite-variance q-Gaussian, and $\mathcal{N}_t$ normalizes the sampled time grid. Equation~\eqref{eq:ig} is recovered exactly when $q=\beta=1$. Depending on the free parameters, the model is called q-FPT, time-FPT, or memory-qFPT. Equation~\eqref{eq:qfpt} is only a reduced observation model. It is not claimed to be an exact solution of the nonlinear fractional absorbing-boundary problem.

Two models with explicit transport structure are also tested. In Gamma-CTRW, a mobile ADE fraction is combined with an ADE first-passage density that is convolved with a Gamma waiting-time kernel:
\begin{equation}
f_{\mathrm{GCTRW}}(t)=w f_{\mathrm{ADE}}(t)
+(1-w)\,[f_{\mathrm{ADE}}*g(k,\theta)](t).
\label{eq:ctrw}
\end{equation}
The finite-mean kernel is not proposed as a universal CTRW law. It is used to test whether an explicit residence-time state is needed. The two-path model is
\begin{equation}
f_{\mathrm{2path}}(t)=w f_{\mathrm{IG}}(t;D_1,v_1)
+(1-w)f_{\mathrm{IG}}(t;D_2,v_2),
\label{eq:mixture}
\end{equation}
This model tests whether separate fast and slow pathways are present. The six models are summarized in Table~\ref{tab:models}.

\begin{table}[htbp]
\centering
\caption{First-passage-aware models. The parameter count $k$ is used in AICc.}
\label{tab:models}
\begin{tabularx}{\textwidth}{lclY}
\toprule
Model & $k$ & Free structural parameters & Interpretation tested \\
\midrule
ADE-FPT & 2 & $D,v$ & Classical drift-dispersion first passage \\
q-FPT & 3 & $q,D,v$ & Non-Gaussian spatial shape \\
time-FPT & 3 & $\beta,D,v$ & Single fractional-memory exponent \\
memory-qFPT & 4 & $q,\beta,D,v$ & Joint nonlinearity and memory hypothesis \\
Gamma-CTRW & 5 & $D,v,k,\theta,w$ & Explicit finite-mean waiting-time state \\
two-path-IG & 5 & $D_1,v_1,D_2,v_2,w$ & Coexisting slow and fast pathways \\
\bottomrule
\end{tabularx}
\end{table}

\section{Methods}

\subsection{Numerical verification and high-q audit}

A mass-conservative finite-volume scheme is used for the spatial PDE. Advection is treated by upwinding, and nonlinear diffusion is linearized by Picard iteration. Backward Euler is used at $\beta=1$; otherwise, an L1 Caputo history operator is applied. Nine ADE benchmarks are formed from $D=0.001,0.01,0.1$ and $v=0,0.5,1$. Relative $L_2$ error, mass error, nonnegativity, and peak position are checked against the analytical Gaussian solution.

For the high-$q$ test, $q=1.35,1.50,1.60$ is used on domains with half-widths 6, 12, and 24. A q-Gaussian initial condition is imposed so that an arbitrary pulse shape is avoided. Exponents are estimated with shifted power laws. Same-model inverse tests are then run for 100 parameter sets, five observation distances, and 1-10\% multiplicative noise. The numerical and inference codes are verified by these tests. No evidence for aquifer physics is assigned to them.

\subsection{Digital aquifers and simulated data}

Four conductivity classes, H0-H3, are generated. Their target $\operatorname{Var}(\ln K)$ values are 0, 0.5, 1.5, and 3.0. Three independent realizations are used per class. Connected high-conductivity channels and low-conductivity retention patches are added to H3. Steady flow is solved on a $96\times48$ grid at three flow levels. In each of the 36 runs, 8000 particles are tracked. First crossings are recorded at $x/L=0.2,0.4,0.6,0.8,0.99$. A separate synthetic field is also generated with $\operatorname{Var}(\ln K)=3.5$, new correlation scales and seed, and 14000 particles.

Particles are moved through the two-dimensional velocity field by a random-walk scheme. Transverse velocity and dispersion are included, and waiting is added in low-velocity cells. In total, 185 digital BTCs are obtained. They provide a controlled coarse-graining test, not observations from a physical aquifer. Laboratory-like records were also generated in the wider project. Instrument response, noise, detection limits, and mass-recovery variation were added. These records are not used to support physical validation.

\subsection{Inference and falsification tests}

All five distances are fitted jointly. Log-density residuals above the detection threshold are minimized. Global differential evolution is followed by bounded nonlinear least squares. Parameter count is penalized with corrected Akaike information criterion (AICc) \citep{burnham2002}. A model is labeled the AICc winner when its $\Delta$AICc is zero within numerical tolerance.

In the main extrapolation test, $x/L=0.2,0.4,0.6$ are fitted and 0.8 and 0.99 are predicted without refitting. Two more transfer tests are applied to memory-qFPT. Parameters from flow levels V1-V2 are used to predict V3, and parameters from realizations R1-R2 are used to predict R3. Next, $\alpha$ is calculated from fitted $q,\beta$ with Eq.~\eqref{eq:scaling}. It is compared with the exponent measured independently from particle-plume variance. The external synthetic field is tested with the same three-distance/two-distance split.

Multimodality is measured at $x/L=0.99$. Peaks must have a prominence greater than 5\% of the maximum BTC. The test shows whether the shape of the arrival distribution can be represented by one self-similar state.

\begin{figure}[htbp]
\centering
\includegraphics[width=\textwidth]{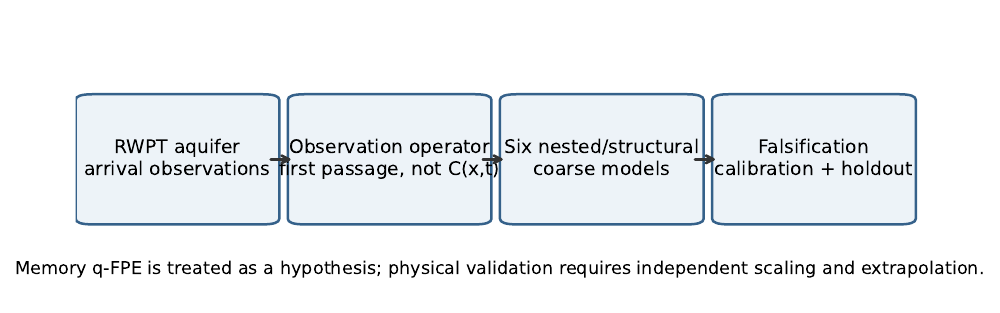}
\caption{Verification-to-falsification workflow. Observation consistency is imposed before model selection; independent scaling and holdout prediction remain necessary after correction.}
\label{fig:workflow}
\end{figure}

\section{Results}

\subsection{Verification resolves the apparent high-q bias}

All nine ADE benchmarks are passed. The maximum relative $L_2$ error is 0.00367, and the maximum relative mass error is $8.31\times10^{-13}$. Peak displacement never exceeds one grid cell. The MSD exponent remains sensitive to domain size at high $q$. For $q=1.60$, its relative error falls from 12.7\% to 6.45\% when the domain half-width is increased from 6 to 24. On the same domains, IQR$^2$ errors are 1.53\%, 1.12\%, and 1.12\%. The maximum error on the largest domains is therefore 1.12\%. The earlier high-$q$ discrepancy is explained by tail truncation and the chosen estimator, not by failure of the similarity law.

Accurate recovery is obtained when the same model is used to generate and fit the data. Median absolute errors over 100 cases are 0.000668 for $q$, 0.000344 for $\beta$, 0.000766 for $\log_{10}D$, and 0.000474 for $v$. When several distances are used, the median $\beta$ error is reduced by a factor of 16.7. Numerical consistency and matched-model identifiability are therefore established. Uniqueness under model error is not.

\subsection{First-passage correction reverses the model narrative}

For the homogeneous H0 cases, the normalized spatial propagator is replaced by the exact ADE first-passage density. Median log-RMSE is reduced from 1.019 to 0.849, an improvement of 16.7\%. Thus, part of the earlier support for anomalous parameters was created by the observation operator.

\begin{figure}[htbp]
\centering
\includegraphics[width=\textwidth]{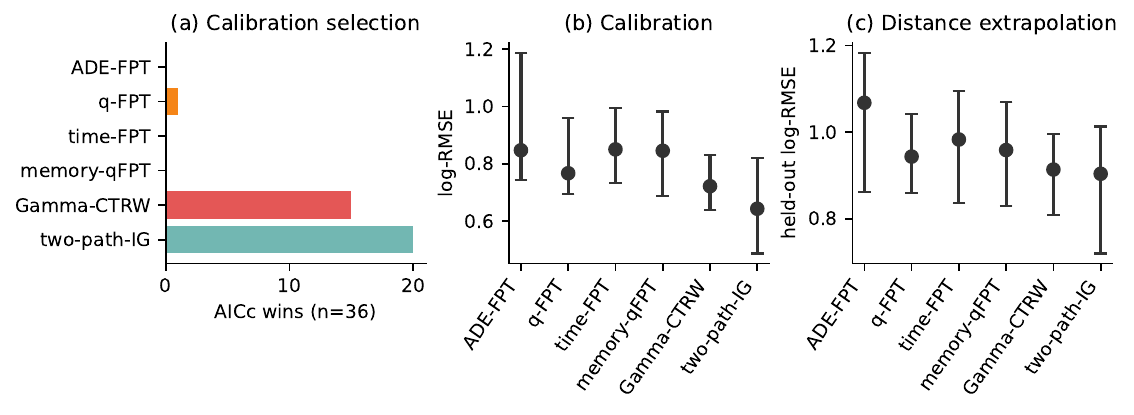}
\caption{First-passage-aware model comparison across 36 digital-aquifer runs. Points in panels (b,c) are medians and bars span the interquartile range.}
\label{fig:performance}
\end{figure}

The full comparison gives a clear result (Fig.~\ref{fig:performance}; Table~\ref{tab:performance}). Two-path-IG is selected by AICc in 20 of 36 runs. Gamma-CTRW is selected in 15, and q-FPT is selected in one. ADE-FPT, time-FPT, and memory-qFPT are never selected. For memory-qFPT, the median $\Delta$AICc is 152.5, with a range of 31.3-304.4. Its lack of wins is therefore not caused by near ties. The lowest median calibration log-RMSE is obtained by two-path-IG (0.643). Gamma-CTRW (0.722) and q-FPT (0.767) follow. Despite two extra parameters, memory-qFPT (0.845) is almost identical to ADE-FPT (0.847).

\begin{table}[htbp]
\centering
\caption{Performance across 36 digital-aquifer runs. Lower errors are better.}
\label{tab:performance}
\resizebox{\textwidth}{!}{%
\begin{tabular}{lrrr}
\toprule
Model & AICc wins & Median calibration log-RMSE & Median distance-holdout log-RMSE \\
\midrule
ADE-FPT & 0 & 0.847 & 1.068 \\
q-FPT & 1 & 0.767 & 0.943 \\
time-FPT & 0 & 0.850 & 0.983 \\
memory-qFPT & 0 & 0.845 & 0.959 \\
Gamma-CTRW & 15 & 0.722 & 0.914 \\
two-path-IG & 20 & 0.643 & 0.904 \\
\bottomrule
\end{tabular}}
\end{table}

\begin{figure}[htbp]
\centering
\includegraphics[width=\textwidth]{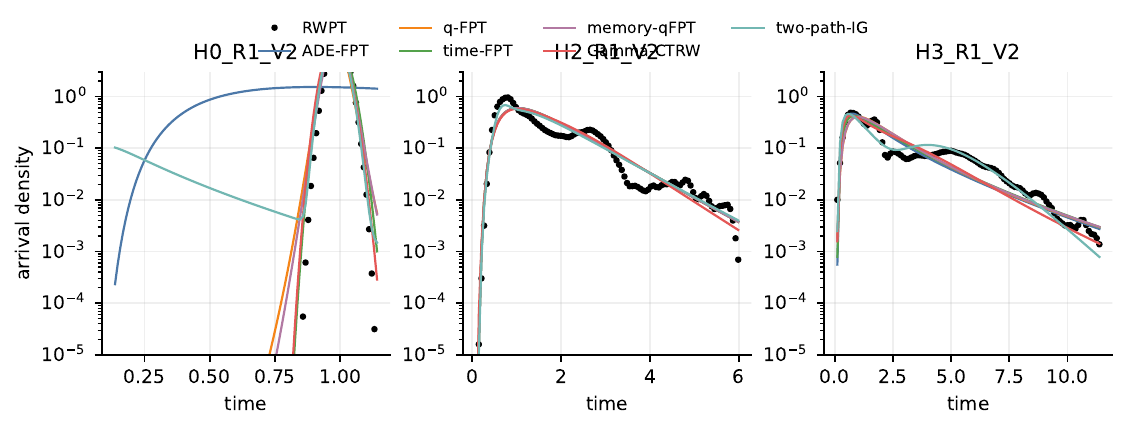}
\caption{Observed and fitted downstream BTCs at $x/L=0.99$. The logarithmic ordinate emphasizes late arrivals. Lines are calibration fits, not holdout predictions.}
\label{fig:btcs}
\end{figure}

The reason for this ranking is seen in representative BTCs (Fig.~\ref{fig:btcs}). H0 is close to one narrow first-passage peak. In H2 and H3, shoulders, secondary peaks, and late-time structure are formed. Width and tail weight can be changed by a single q-Gaussian state, but separate modes cannot be produced. Those modes can be represented by the two-path and waiting-time models.

\subsection{Extrapolation and independent scaling reject physical validation}

The same general ranking is retained in the distance holdout. The lowest median log-RMSE is obtained by two-path-IG (0.904), followed by Gamma-CTRW (0.914). Values of 0.943 and 0.959 are obtained for q-FPT and memory-qFPT. Only a small gain over ADE-FPT is provided by memory-qFPT. For flow transfer, its median log-RMSE is 0.867. For transfer to an unseen realization, the error rises to 1.493. Greater sensitivity is therefore shown to unresolved spatial structure than to the imposed mean flow.

\begin{figure}[htbp]
\centering
\includegraphics[width=0.96\textwidth]{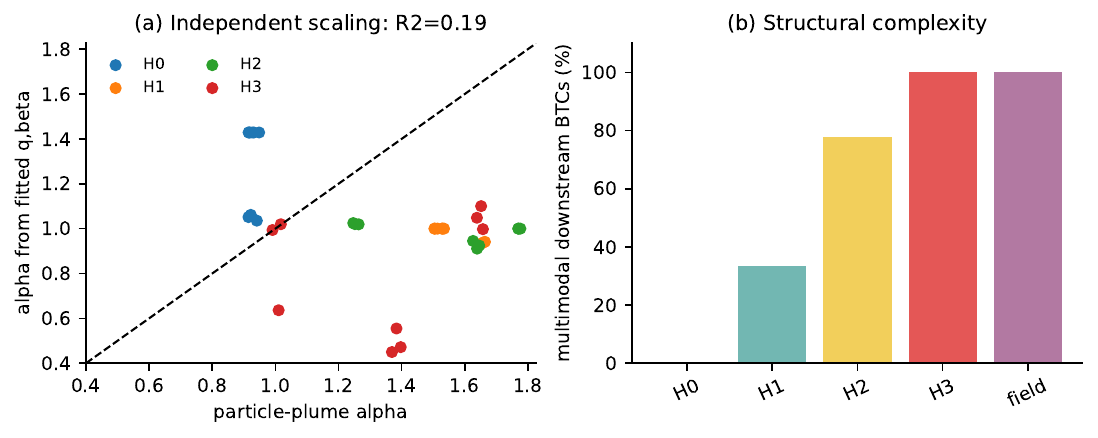}
\caption{Independent physical diagnostics. (a) The spreading exponent predicted from memory-qFPT parameters does not follow the one-to-one line. (b) Multimodality increases systematically with heterogeneity.}
\label{fig:scaling}
\end{figure}

The independent scaling test gives a stronger rejection. The fitted parameters give a regression slope of $-0.332$, an intercept of 1.466, and $R^2=0.189$ against the particle-plume exponent (Fig.~\ref{fig:scaling}a). The median absolute error is 0.521. Although the association is nonzero ($p=0.008$), its sign is wrong and its predictive value is low. Statistical significance cannot be treated as physical validation when the expected one-to-one relation is not recovered.

This conclusion is reinforced by parameter-bound hits. Among 36 memory-qFPT fits, $q$ is within 1\% of its lower bound in 13 cases and its upper bound in 8. The upper $\beta$ bound is reached in 19 cases, and the lower bound is reached in 3. Weak or compensating identification is indicated. A correlation is found between $\beta$ and several heterogeneity measures. For example, Spearman $\rho=-0.638$ is obtained for $\operatorname{Var}(\ln K)$. This descriptive link does not offset the failed scaling and model-selection tests.

\subsection{Multimodality and the external synthetic field identify missing structure}

No H0 BTC is multimodal at the downstream plane. The fraction rises to 33.3\% for H1, 77.8\% for H2, and 100\% for H3 (Fig.~\ref{fig:scaling}b). Multimodality is also found in the independent synthetic field. This steady increase explains why support for a one-state self-similar model is lost as heterogeneity grows.

\begin{figure}[htbp]
\centering
\includegraphics[width=\textwidth]{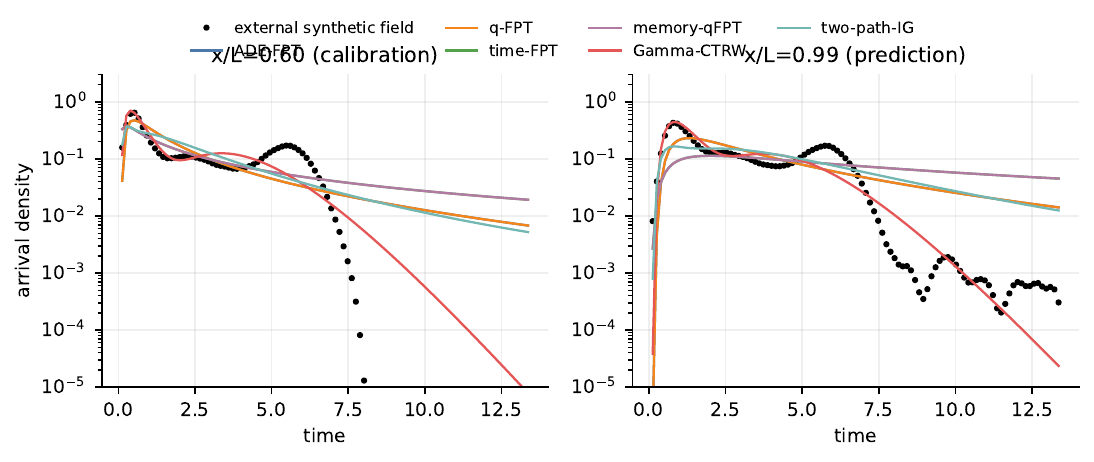}
\caption{External synthetic field under a three-distance calibration and two-distance holdout. The downstream panel is a prediction without refitting.}
\label{fig:field}
\end{figure}

For the external synthetic field, a calibration log-RMSE of 0.846 is obtained by Gamma-CTRW, which is selected by AICc. The next-best value is 1.296 for two-path-IG. The other four models all give about 1.382. In the distance holdout, Gamma-CTRW predicts the two downstream planes with log-RMSE 1.100. The errors are 2.120 for two-path-IG, 2.201 for ADE-FPT/q-FPT, and 2.585 for memory-qFPT. The calibrated $x/L=0.60$ curve and the unrefitted $x/L=0.99$ prediction are shown in Fig.~\ref{fig:field}. The field is simulated, so empirical validation is not provided. However, a real generator shift is present in correlation scale, variance, seed, and particle count.

\section{Discussion}

\subsection{Observation consistency is necessary but not sufficient}

The main observation error is removed by the first-passage correction. The homogeneous benchmark is improved by Eq.~\eqref{eq:ig}. In addition, every nested anomalous model is reduced to the correct ADE observation law by Eq.~\eqref{eq:qfpt}. A missing $x/t$ factor can no longer be mistaken for anomalous dynamics.

Memory q-FPE is not validated by this correction. After models with explicit structure are included, no AICc comparison is won by memory-qFPT. Calibration is weaker than for q-FPT alone. The independent scaling law is missed, and extrapolation is less accurate than for the pathway and waiting-time models. This negative result is informative. The earlier memory parameters were partly used to absorb both an operator error and missing transport states.

Parameter values could change if an exact nonlinear fractional first-passage solver were used. Even then, three independent features would still need to be reproduced: multimodality, plume spreading, and holdout behavior. These features were not part of the calibration target. They are therefore not artifacts of the reduced observation model. The next test should use an absorbing boundary and a mass-flux observation. It should also retain the same structural alternatives and holdout design.

\subsection{What the preferred comparators imply}

Selection of two-path-IG in 20 runs is consistent with fast connected channels and slower background flow. Gamma-CTRW is selected in 15 runs and is strongly favored in the external synthetic field. Thus, an explicit residence-time state appears useful when delay is distributed rather than split into two peaks. Neither model is claimed to be the unique upscaled equation. Instead, they show that transport structure missing from a one-state memory q-FPE can be detected in the same BTCs.

The role of $q$ is also clarified. Lower median calibration and holdout errors are obtained by q-FPT than by memory-qFPT, and one AICc comparison is won. Heavy spatial tails may therefore be useful in some runs. The added fractional-memory exponent is not independently supported. The boundary hits and failed $\alpha$ relation instead show compensation between $q$ and $\beta$. They cannot yet be treated as transferable aquifer properties.

\subsection{Implications for upscaling studies}

Three practices are suggested. First, the observation operator should be derived from the measurement. First-crossing flux is needed for arrival records, resident concentration for occupancy measurements, and flux concentration for mass discharge. Second, numerical verification, same-model recovery, and physical validation should be reported separately. Third, a model should be tested against alternatives with different mechanisms, not only against nested limits. A small gain over ADE is not sufficient when held-out distances are predicted much better by a pathway or waiting-time model.

A numerical caution is added by the high-$q$ audit. Near the finite-variance limit, a second-moment estimate can fail because the domain is too small. The PDE exponent need not be wrong. Quantile widths and larger domains provide simple checks. At the same time, correct scaling in the ideal equation does not ensure that fitted parameters recover plume scaling in a heterogeneous aquifer.

\subsection{Limitations and required empirical tests}

All aquifer, laboratory-like, and field-like data are simulated. The RWPT generator includes several forms of heterogeneity, but important processes are omitted. These include reactive chemistry, density effects, unsaturated flow, three-dimensional connectivity, and uncertain sampling volumes. Only a finite-mean Gamma kernel is used for CTRW. The two-path mixture is descriptive and may simplify a broad velocity distribution too strongly.

Equation~\eqref{eq:qfpt} remains a reduced first-passage model. For a definitive PDE test, the nonlinear time-fractional equation must be solved with an absorbing boundary and a mass-flux observation. Conservation of surviving mass and convergence of the exit-time density must then be checked. Physical validation also requires preregistered column or sandbox experiments and an independent field tracer test. Several distances should be sampled. Early arrivals must be resolved, and the late-time range should span several orders of magnitude. Replicate realizations or wells are also needed. These requirements are not replaced by the synthetic external field used here.

\section{Conclusions}

The BTC observation problem has been corrected at the reduced-model level. Five conclusions are supported:
\begin{enumerate}
  \item High-$q$ scaling is numerically consistent when finite-domain tail bias is controlled with IQR$^2$; the largest-domain exponent error is at most 1.12\%.
  \item Exact ADE first passage improves the homogeneous benchmark, confirming that normalized spatial propagators should not be used as arrival-time likelihoods.
  \item Memory-qFPT is not selected in any of 36 corrected AICc comparisons and does not have the best median distance extrapolation.
  \item Its fitted $q,\beta$ fail the independent plume-scaling relation, so the parameters cannot presently be interpreted as validated aquifer properties.
  \item The rise of multimodality and the success of two-path and Gamma-CTRW alternatives point to unresolved pathway and waiting-time structure.
\end{enumerate}
Memory q-FPE should therefore remain a testable hypothesis. Its physical validation claim is rejected under the present synthetic design. This result does not show that every absorbing-boundary or multistate extension must fail.

\section*{Data and code availability}

All data used in this study are program-generated simulated data. Processed BTCs, fitted parameters, holdout results, figure-ready tables, source code, environment specification, and a SHA-256 manifest are supplied with this submission. No confidential or third-party data are used. The complete package will be deposited in a public repository upon acceptance.

\section*{CRediT authorship contribution statement}

Ke-Ming Shen: Conceptualization, Methodology, Software, Formal analysis, Investigation, Visualization, Writing - original draft, Supervision, Funding acquisition. Yi-Xuan Liu: Methodology, Validation, Formal analysis, Data curation, Writing - review and editing.

\section*{Declaration of competing interest}

The authors declare that they have no known competing financial interests or personal relationships that could have appeared to influence the work reported in this paper.

\section*{Funding}

This work is supported in part by Innovative Platform for Quark Matter Science and Advanced Detection Technology funded under Ministry of Education's ``Two-New'' Initiative (No. QLPL2026P01).

\section*{Declaration of generative AI and AI-assisted technologies}

During preparation of this manuscript, the authors used OpenAI Codex to assist with code organization, language editing, and preparation of submission files. All numerical methods, results, citations, interpretations, and final text were reviewed and verified by the authors, who take full responsibility for the content.

\section*{Acknowledgments}

The authors thank Yang Liu and Siqi Yin for insightful discussions and technical support.

\bibliographystyle{plainnat}
\bibliography{references}

\end{document}